# Model of Transcriptional Activation by MarA in *Escherichia coli*

Michael E. Wall[1,2,3,*], David A. Markowitz[1], Judah L. Rosner[4], Robert G. Martin[4]

[1] Computer, Computational, and Statistical Sciences Division,

[2] Bioscience Division, and

[3] Center for Nonlinear Studies, Los Alamos National Laboratory, Los Alamos, NM, USA

[4] Laboratory of Molecular Biology, National Institute of Diabetes, Digestive and Kidney Diseases, National Institutes of Health, Bethesda, MD, USA

[*]**Corresponding Author**:

Michael E. Wall

CCS-3, Mail Stop B256

Los Alamos National Laboratory

Los Alamos, NM, 87545

*Tel*: (505)665-4209

*Fax*: (505)667-1126

*E-Mail*: mewall@lanl.gov

**Major Category**: Biological Sciences

**Minor Category**: Biophysics and Computational Biology

**Manuscript length**: 22 text pages, 2 figures, and 2 tables

**Abbreviations**: CRP, cAMP receptor protein; IPTG, isopropyl β-D-1-thiogalactopyranoside

**Technical Release Number**: LA-UR-09-00238

## Abstract

We have developed a mathematical model of transcriptional activation by MarA in *Escherichia coli*, and used the model to analyze measurements of MarA-dependent activity of the *marRAB, sodA*, and *micF* promoters in *mar-rob-* cells. The model rationalizes an unexpected poor correlation between the mid-point of *in vivo* promoter activity profiles and *in vitro* equilibrium constants for MarA binding to promoter sequences. Analysis of the promoter activity data using the model yielded the following predictions regarding activation mechanisms: (1) MarA activation of the *marRAB*, *sodA,* and *micF* promoters involves a net acceleration of the kinetics of transitions after RNA polymerase binding, up to and including promoter escape and message elongation; (2) RNA polymerase binds to these promoters with nearly unit occupancy in the absence of MarA, making recruitment of polymerase an insignificant factor in activation of these promoters; and (3) instead of recruitment, activation of the *micF* promoter might involve a *repulsion* of polymerase combined with a large acceleration of the kinetics of polymerase activity. These predictions are consistent with published chromatin immunoprecipitation assays of interactions between polymerase and the *E. coli* chromosome. A lack of recruitment in transcriptional activation represents an exception to the textbook description of activation of bacterial $\sigma^{70}$ promoters. However, use of accelerated polymerase kinetics instead of recruitment might confer a competitive advantage to *E. coli* by decreasing latency in gene regulation.

\body

**Introduction**

In the textbook model of transcriptional activation, activator recruits RNA polymerase to the promoter (1-3). Recruitment works purely by increasing the likelihood that there is an enzyme present, poised to synthesize mRNA; it does not require changes in the downstream events that lead to transcription. This scenario currently dominates interpretation of bacterial transcriptional activation data, and has been adopted in statistical-thermodynamic models of transcriptional activation that capture the activator-dependent expression of several promoters in *Escherichia coli* (4, 5).

The simplicity and generality of the recruitment model are appealing; however, mechanisms of transcriptional activation can vary depending on the activator (3), and are sometimes surprising. For example, MerR binds between the -10 and -35 RNA polymerase recognition sequences (6), and might be expected to repress transcription by blocking polymerase interactions with the promoter. Instead of blocking polymerase, however, MerR and related transcription factors such as SoxR (7) activate transcription by extending the DNA between the hexamers to a length more appropriate for open complex formation.

In addition to diversity among activators, a single activator can work differently at different promoters. For example, cAMP receptor protein (CRP) binds at different locations upstream of the *gal* (-41.5), *lac* (-61.5), and *malT* (-70.5) promoters in *E. coli*, with diverse consequences for activation (8). Interactions between CRP and polymerase vary for *gal* or *lac*, and activation at *malT* can involve an accelerated escape of polymerase from the initiation complex, without detectable changes in the events leading up to open complex formation (8, 9).

Like CRP, MarA activates transcription from diverse locations upstream of promoters; in addition, it binds to an asymmetric recognition motif with opposite orientations depending on the location (10, 11). However, unlike CRP, which is controlled by intracellular cAMP, MarA has only one domain that interacts with both DNA and polymerase, and has no known effectors. In addition, MarA is a functional monomer whereas most well-characterized transcription factors, like CRP, are known to function as a dimer or a higher order complex.

Because the position and orientation of its recognition motifs are diverse, and its interactions are relatively simple, MarA is an ideal system for studying how activators differentially activate transcription from promoters. In addition, like CRP, MarA is a global regulator: it activates ~40 genes (the *marA/soxS/rob* regulon) of the *E. coli* chromosome resulting in different levels of resistance to a wide array of antibiotics and superoxides (see (12) for references). Diversity in transcriptional activation by MarA therefore presumably has important functional consequences for *E. coli*.

To characterize diversity in MarA regulation of promoter activity, we placed the expression of MarA under the control of the LacI repressor, determined the relationship between isopropyl β-D-1-thiogalactopyranoside (IPTG) concentration and the intracellular concentration of MarA, and examined the expression of 10 promoters of the regulon as a function of activator concentration (13). We found that: (i) the MarA concentrations needed for half-maximal activation varied by at least 19-fold among the promoters indicating substantial variation in promoter activities; (ii) only *marRAB*, *micF*, and *sodA* were saturated at the highest level of MarA obtained; and (iii) the correlation between the MarA concentration needed for half-maximal promoter activity *in vivo* and *marbox* binding affinity *in vitro* was poor.

To understand the source of the diversity in MarA activation of promoters, we develop here a quantitative model of MarA transcriptional activation of *marRAB*, *sodA*, and *micF*—the only promoters that exhibited saturation at the highest levels of MarA (13). Our model uses a statistical-thermodynamic treatment of promoter states (14), and considers the interaction between activator and polymerase away from the promoter, which, to our knowledge, has not been considered in previous gene regulation models. The model rationalizes the poor correlation between *in vivo* promoter activity profiles and *in vitro* activator binding affinities. It also suggests that there are diverse mechanisms of MarA activation for the *marRAB*, *sodA*, and *micF* promoters.

## Model

We considered a statistical-thermodynamic model of promoter states that was originally developed to study transcriptional repression by λ phage repressor (14). The model is illustrated in Fig. 1. In State 0, the promoter is free. This is the reference state with energy $\Delta G_0 = 0$. In State A, MarA is bound at the operator sequence $O_A$, yielding free energy $\Delta G_A$; in State R, polymerase is bound at the promoter $P$, yielding free energy $\Delta G_R$; and in State X, both MarA and polymerase are bound, yielding free energy $\Delta G_X$. These free energies are defined for 1 M concentrations of "free" MarA ($\Delta G_A$), polymerase ($\Delta G_R$), and MarA-polymerase complex ($\Delta G_X$). (We use a liberal definition of free molecules in which they may be located anywhere away from the promoter, including nonspecific sites on DNA, and use a single effective free energy to characterize the equilibrium with the bound state.)

The free energies of the states are related to corresponding dissociation constants via $\Delta G_i = k_B T \ln K_i$, where $K_i$ is the dissociation constant of state $i$ in molar units. These dissociation constants in turn determine the statistical weights $p_i$ via the following equations:

$$\begin{aligned}
p_A &= \frac{[A]}{K_A} p_0, \\
p_R &= \frac{[R]}{K_R} p_0, \\
p_X &= \frac{[X]}{K_X} p_0, \text{ and} \\
p_0 &= \frac{1}{1 + [A]/K_A + [R]/K_R + [X]/K_X}.
\end{aligned} \qquad (1)$$

In Eqs. (1), the first three equations follow from the definition of the dissociation constants, and the last equation follows from the normalization condition $\sum_i p_i = 1$. Consistent with the definitions of free energies in the previous paragraph, the ratios $p_i/p_0$ for 1 M concentrations of the DNA-binding partner are equal to the Boltzmann factor $e^{-\Delta G_i/k_B T}$.

The equilibrium between free MarA (A) and polymerase (R) and the MarA-polymerase complex (X) is modeled assuming steady-state equilibration characterized by dissociation constant $K_{AR}$. We assume that interactions with the promoter do not significantly influence the equilibrium. This is a reasonable assumption given that the chromosomal *lacZ* reporter fusions used in Ref. (13) have a copy number of at most 5 per cell. The model leads to the following equations

$$\begin{aligned}
K_{AR} &= [A][R]/[X], \\
[R]_T &= [R] + [X], \text{ and} \\
[A]_T &= [A] + [X],
\end{aligned} \qquad (2)$$

where $[R]_T$ and $[A]_T$ are the total levels of polymerase and MarA in the cell, respectively. The solution of Eqs. (2) is

$$\begin{aligned}
[A] &= \frac{1}{2}\left[\sqrt{(K_{AR} - [A]_T + [R]_T)^2 + 4[A]_T K_{AR}} - (K_{AR} - [A]_T + [R]_T)\right], \\
[R] &= \frac{1}{2}\left[\sqrt{(K_{AR} + [A]_T - [R]_T)^2 + 4[R]_T K_{AR}} - (K_{AR} + [A]_T - [R]_T)\right], \text{ and} \\
[X] &= [A][R]/K_{AR}.
\end{aligned} \qquad (3)$$

For given values of $[R]_T$ and $[A]_T$, Eqs. (4) yield the concentrations that determine the state weights in Eqs. (1).

The total promoter activity is a weighted sum of the activities in each state. No transcription occurs in states 0 or A, in which polymerase is absent from the promoter. Transcription occurs in state R with activity $a_R$, and in state X with activity $a_X$; polymerase is present at the promoter in both of these states. The equation for the total activity $a$ is

$$a = a_R p_R + a_X p_X. \qquad (4)$$

We use Eq. (2) to model assays of β-galactosidase activity (Methods) (13); in doing this, we assume that the underlying promoter activity is proportional to the measured β-galactosidase activity resulting from *lacZ* reporter expression. The activity assays were performed after many generations and are assumed to represent steady-state levels.

## Results

***Calibration of IPTG against MarA***. We calibrated IPTG levels against MarA levels using analyses of Western blots in multiple lanes from a single gel (13). The MarA vs. IPTG data are well-described using the equation

$$[A]_T = [A]_{max} \frac{[I]^h}{[I]^h + K_I^h}, \qquad (5)$$

where $[I]$ is the extracellular IPTG concentration, $[A]_T$ is the total cellular MarA concentration that appears in Eqs. (3), $[A]_{max}$ = 20,983 molecules cell$^{-1}$, $K_I$ = 20.132 mM, and $h$ = 2.576 (Supporting Information Fig. S1). In the absence of IPTG, cells contained a small amount of MarA that decreased in cells carrying a control plasmid. However, when we added this basal level of MarA to Eq. (5), we found that we were unable to explain the sensitivity of promoter activity to low levels of IPTG. In addition, all data points in the gel, except for a measurement at 2 uM IPTG, were consistent with the absence of MarA at low IPTG. We therefore used the simpler form of Eq. (5) for the modeling.

***The model is consistent with promoter activation data***. The best-fit activation profiles for each promoter are illustrated in Fig. 2; the quality of the fits indicates that the model is entirely consistent with the observed IPTG-dependent activation of the *marRAB*, *sodA* and *micF* promoters.

***MarA increases polymerase activity***. To determine whether polymerase activity changes when MarA is bound to the promoter, we calculated the ratio $a_X/a_R$ for all models. For each model with each set of parameter values, both $a_X$ and $a_R$ were obtained using Eq. (2) to perform a linear regression (Model). Results are expressed in terms of the acceleration energy, $e_a$, defined as

$$e_a = -k_B T \ln \frac{a_X}{a_R}. \tag{6}$$

The acceleration energy is equivalent to the activator-induced change in activation energy of a lumped transcription initiation process, under the assumption that $a_X$ and $a_R$ each follow an Arrhenius law with the same attack frequency. A value $e_a = 0$ corresponds to $a_X = a_R$; this condition is consistent with a strict recruitment model of transcriptional activation, in which activator increases the occupancy of polymerase at the promoter but does not alter the kinetics of polymerase activity (1, 3). Models with $e_a < 0$ exhibit acceleration and models with $e_a > 0$ exhibit retardation of polymerase activity in the presence of activator.

The acceleration energy is negative for all 10,000 sets of parameter values in each promoter activation model (Supporting Information Fig. S2, left panels). Activator therefore increases polymerase activity in all promoter activation models. For each promoter, the value of $e_a$ corresponding to the minimum in $\chi^2$ is in the neighborhood of $k_B T$ (Table II).

***MarA does not recruit polymerase to the* marRAB *and* sodA *promoters*.** To determine whether the affinity of polymerase for the promoter changes in the presence or absence of MarA, we calculated the ratio between corresponding polymerase-DNA dissociation constants. The dissociation constant in the absence of MarA is just $K_R$. The dissociation constant in the presence of MarA, $K_R^+$, is determined by dissociation constants given in Table I using detailed balance (Fig. 1, right panel):

$$K_R^+ = \frac{K_X K_{AR}}{K_A}. \tag{7}$$

We calculated the ratio $K_R^+/K_R$ and used the recruitment energy, $e_r$, to characterize the change in polymerase-DNA affinity upon MarA binding:

$$e_r = k_B T \ln \frac{K_X K_{AR}}{K_A K_R} = \Delta G_X + k_B T \ln K_{AR} - \Delta G_A - \Delta G_R \tag{8}$$

.

From the definition of $\Delta G_X$ (Model), $\Delta G_X + k_B T \ln K_{AR}$ is equal to the free energy of the MarA-polymerase-DNA complex in the presence of 1 M free MarA and 1 M free polymerase. Therefore, from the definitions of $\Delta G_A$ and $\Delta G_R$ (Model) the recruitment energy $e_r$ is equal to the free energy of interaction between MarA and polymerase on the DNA in the presence of 1 M each free MarA and free polymerase. A value $e_r = 0$ indicates no interaction between MarA and polymerase; a value $e_r < 0$ indicates that MarA attracts polymerase to the promoter; and a value $e_r > 0$ indicates that MarA repels polymerase from the promoter. For the parameter ranges in Table I, $e_r$ assumes either positive or negative values (Supporting Information Fig. S2, right panels).

For *marRAB*, the models with the lowest $\chi^2$ have $e_r < 0$ (Table II; Supporting Information Fig. 2a, 2b, right panels;). MarA activation of *marRAB* therefore involves attraction of polymerase to the promoter by MarA. For *sodA*, the models with the lowest $\chi^2$ have $e_r$ slightly less than 0, indicating that MarA weakly attracts polymerase to the promoter. We also compared the total occupancy of polymerase at the promoter,

$$p_{RX} = p_R + p_X, \tag{9}$$

in the presence ($p_{RX}^+$) vs. the absence ($p_{RX}^-$) of MarA. For both *marRAB* and *sodA*, the lowest-$\chi^2$ models showed both a basal occupancy $p_{RX}^-$ and ratio $p_{RX}^+/p_{RX}^-$ equal to 1 (Table II, Supporting Information Fig. S3). MarA therefore does not recruit polymerase to these promoters.

***MarA repels polymerase from the* micF *promoter***. For *micF*, the models with the lowest $\chi^2$ have $e_r > 0$ (Table II; Supporting Information Fig. S2c, right panel; Table II). Activation in this model therefore involves *repulsion* of polymerase from the promoter by MarA. Moreover, unlike the neutral effect of attraction for *marRAB* and *sodA*, analysis of $p_{RX}^+/p_{RX}^-$ and $p_{RX}^-$ indicates that repulsion leads to a decrease in the occupancy of polymerase at *micF* in the presence vs. the absence of MarA (Table II; Supporting Information Fig. S3).

***Results are robust to parameter variation***. To quantify the degree of uncertainty in estimated parameter values within the nominal range, we calculated asymmetric errors of parameter values with respect to the optimum (Table II). The squared errors for parameter *x* were calculated using the equation

$$\begin{aligned}\sigma_{+x}^2 &= \sum_{i:x_i > x_{\min}}(x_i - x_{\min})^2 e^{-\chi_i^2/2} \Big/ \sum_{i:x_i > x_{\min}} e^{-\chi_i^2/2} \\ \sigma_{-x}^2 &= \sum_{i:x_i < x_{\min}}(x_i - x_{\min})^2 e^{-\chi_i^2/2} \Big/ \sum_{i:x_i < x_{\min}} e^{-\chi_i^2/2}\end{aligned}, \tag{9}$$

Where $x_i$ is the value of parameter $x$ in the $i^{th}$ sample, $x_{min}$ is the value of $x$ in the sample with the lowest value of $\chi^2$, and $\chi_i^2$ is the value of $\chi^2$ for the $i^{th}$ sample. In using the likelihood function $e^{-\chi_i^2/2}$, we assume that the errors in measurements of mean promoter activity are independent and normally distributed with widths equal to the standard error of the mean (error bars in Fig. 2 and error values in Supporting Information Table S1). The values of $a_R$, $e_a$, $p_{RX}^-$, and $p_{RX}^+/p_{RX}^-$ in Table II are well-constrained by the data given the nominal ranges in parameter values (Table I). The sign of $e_r$ is positive for *marRAB* and negative for *micF*; for *sodA*, the value is slightly negative, but indistinguishable from 0 given the errors (Table II; Supporting Information Fig. S2). The absolute parameter values of $K_R$ and $K_X$ are poorly constrained (not shown), but their ratio is well-constrained (and is related to $e_r$ through Eq. (8)).

To quantitatively assess confidence in the finding that MarA repels polymerase from the *micF* promoter, we used Bayesian analysis methods to estimate a cumulative posterior probability distribution of $p_{RX}^+/p_{RX}^-$ values given the range parameter values in Table I. To perform the analysis, we used the likelihood function $e^{-\chi_i^2/2}$ in Eq. (9), and assumed a log uniform prior over parameter ranges. The analysis indicated that, given the assumed parameter ranges and prior, there is a 45% chance that $p_{RX}^+/p_{RX}^-$ is smaller than 0.7, a 55% chance that it is smaller than 0.8, and a 72% chance that it is smaller than 0.9 (the full curve is shown in Supporting Information Fig. S4). Based on this analysis, it seems reasonably likely that MarA repulsion of polymerase from the *micF* promoter is significant.

We also examined the sensitivity of the results to wider parameter variation (Model). All wider variations tested yielded at least some promoter activation curves with reasonable values of $\chi^2$. Decreasing $[R]_T$ to 1,000 or 300 copies per cell lowered the value of $e_r/K_BT$ for the model of *sodA* activation to -3.9 with an error of either 0.2 (for 1,000 copies) or 0.3 (for 300 copies), indicating strong attraction as opposed to very weak attraction for 3,000 copies per cell (Table II). Otherwise, the qualitative mechanisms exhibited by the best-fit models remained robust to wider parameter variations: polymerase is bound at the *marRAB*, *sodA*, and *micF* promoters in the absence of MarA; activation of *marRAB* and *sodA* involves acceleration of polymerase kinetics, and activation of *micF* involves acceleration accompanied by repulsion of polymerase from the promoter.

***Further validation of the model using CRP activation data.*** To further validate our promoter activity model, we used it to analyze published data on transcriptional activation of the *lac* operon by cAMP-CRP (4). The cAMP-CRP-dependent relative promoter activity was represented using the expression $y = 1 + 49x/(x + 5)$, where $x$ is the concentration of the active CRP dimer in nM; this expression is consistent with the data published in Ref. (4). Consistent with expectations (3), we found that recruitment could be important for CRP activation of *lac* (Supporting Information Fig. S5). Recruitment was significant when the dissociation constant between polymerase and the promoter, $K_R$, was sufficiently large ($K_R$ in excess of ~ 100 µM), and the recruitment effect increased with increasing $K_R$. By contrast, we were unable to find any good models for which recruitment is a significant factor in MarA activation of *mar, sodA*, or *micF*, even for large values of $K_R$.

## Discussion

The present model of transcriptional regulation is consistent with available data on MarA-dependent promoter activity (13). The model therefore rationalizes the lack of correlation between measured MarA-DNA dissociation constants and level of MarA required for half-maximal promoter activation. In this regard, a critical feature of our model is explicit consideration of polymerase interactions with MarA both in the absence and in the presence of the promoter: Eq. (1) clearly emphasizes the importance of $K_X$, the dissociation constant between DNA and the MarA-polymerase complex, in determining the activation profile.

We expect interactions with polymerase to be similarly important for other activators, such as CRP, which also binds to polymerase away from the promoter (15). Indeed, our model reproduces a measured CRP-dependent activation profile of the *lac* promoter, and, as expected, predicts that recruitment can be significant in *lac* activation. However, we expect interactions with polymerase to be less important when a repressor decreases expression by interfering with polymerase binding at the promoter. Such interference corresponds to very large values of $K_X$ in our model, which increases the importance of $K_A$ in determining the promoter activity profile. The correlation between a repressor's dissociation constant for binding to a recognition sequence and the level of repressor required for half-maximal repression is therefore expected to be high.

The model predicts that recruitment is not a significant factor in MarA activation of the *marRAB, sodA,* and *micF* promoters. For all of these promoters, the model predicts that polymerase is bound with near unit occupancy in the absence of MarA, and that activation occurs through an increase in polymerase activity when MarA is bound. It is important to note that our model was developed using data from *mar-rob-* strains (13), in which the repressor MarR is absent. In wild-type *E. coli,* we do expect polymerase to be bound at the *sodA* and *micF* promoters in the absence of inducers. However, in wild-type *E. coli*, MarR not only blocks polymerase binding but also blocks MarA binding at *marRAB* (16). We therefore do not expect polymerase to bind at the *marRAB* promoter in the absence of inducers that relieve MarR repression.

These predictions are consistent with two genome-wide studies of polymerase interactions with the *E. coli* chromosome (17, 18). Grainger et al. (17) reported detection of polymerase at the *sodA* promoter but not the *marRAB* promoter; that study was inconclusive with respect to interactions at the *micF* promoter, which controls expression of an antisense mRNA transcript. In addition, we cross-referenced the oligonucleotide coordinates in Herring *et al*. (18) to transcriptional start sites annotated in the EcoCyc database (19), and found strong-binding 50 bp oligonucleotides correctly positioned with respect to *sodA* (sequence beginning at 4,098,720 upstream of 4,098,780 start) and *micF* (sequence beginning at 2,311,050 upstream of 2,311,106 start), but not *marA* (only weakly binding sequences near 1,617,117 start). The presence of polymerase at *sodA* and *micF* in uninduced cells clearly represents an exception to the regulated recruitment model of transcriptional activation at $\sigma^{70}$ promoters (1, 3), in which activation occurs solely through increasing the occupancy of polymerase at the promoter.

Because our model allows the values of $a_X$ and $a_R$ to differ, it is slightly more complex than the regulated recruitment model of transcriptional activation. This additional complexity is well-motivated when activator interacts with $\alpha$-CTD of polymerase, as is the case for MarA. In the absence of activator, polymerase needs to disengage from sigma to escape the promoter; however, in the presence of activator, polymerase must sever its interactions with both sigma and activator. Because of this effect, tight interactions between activator and polymerase are expected to retard promoter escape and decrease the value of $a_X$ compared to $a_R$. Our results support this expectation: $e_a$ and $e_r$ values in Table II (and for wider parameter ranges) are anti-correlated, indicating that stronger recruitment is associated with lower ratios $a_X/a_R$.

An unexpected result of our study is that activation of *micF* might involve repulsion of polymerase from the promoter by MarA. In light of the discussion above, it is possible that polymerase normally sits at the *micF* promoter, and that binding of activator accelerates the events up to and including promoter escape and message elongation, which effectively decreases the promoter affinity of polymerase. The possibility of this mechanism follows immediately from differentiating Eq. (4) with respect to $[A]_T$, which yields

$$a' = a_R p'_R + a_X p'_X. \tag{10}$$

Eq. (10) indicates that, for negative $p'_R$, positive $p'_X$, and negative $(p'_R + p'_X)$, $a'$ can be positive for $a_X > -a_R p'_R / p'_X$. As this condition can only hold for $a_X > a_R$, repulsion decreases latency in transcription compared to regulated recruitment, which calls for $a_X = a_R$. In addition, whereas recruitment involves a decrease in the polymerase off rate in the presence of activator, repulsion involves an increase in the polymerase off rate; repulsion can therefore decrease latency in de-activation of promoters. Such decreases in latency might confer a competitive advantage to *E. coli* in an ecological context (20). The mechanisms of repulsion, highlighted for the *micF* promoter, and acceleration, highlighted for the *marRAB*, *sodA*, and *micF* promoters, are general and might be important for activation of other promoters in *E. coli* and beyond.

## Methods

We used a wide range of parameter values to model the MarA-dependent activity of the *marRAB, sodA,* and *micF* promoters. In presenting our results, we focus on the values listed in Table I, and later discuss the sensitivity of these results to variations in parameter values. The values in the table and broader ranges were obtained as follows:

$K_{AR}$. Using a liquid chromatography assay, Martin *et al.* (21) measured a 0.3 μM dissociation constant for MarA-polymerase complex formation in a crystallization buffer. Dangi *et al.* (22)

obtained a value of 21 µM in low-salt conditions using NMR. Because we consider the NMR measurement to be more reliable, we selected a nominal value of 21 µM for $K_{AR}$. However, we are uncertain about the correct value to use *in vivo*, especially considering that we are lumping into $K_{AR}$ the effect of nonspecific interactions of polymerase and MarA with DNA. To account for uncertainty in $K_{AR}$, we explored values of 0.3 µM, 1.0 µM, 10 µM, and 100 µM. We expect the value of $K_{AR}$ to be promoter-independent, and therefore only compare models across promoters using the same value of $K_{AR}$.

$K_A$. The nominal value of 75 nM for the MarA-*mar* promoter dissociation constant was obtained from the gel retardation assay in Martin *et al*. (21). The nominal value of 2,000 nM for *sodA* was chosen to be consistent with the lack of binding observed in Martin *et al*. (21). The nominal value of 50 nM for *micF* was chosen from a range of measured values from 8 nM to 80 nM, depending on the preparation (R.G. Martin, unpublished results). To determine whether the qualitative conclusions about activation mechanisms are sensitive to the particular value of $K_A$, we analyzed the model using a wide range of values, 0.25-2,500 nM.

$K_R$. The value of the effective dissociation constant for polymerase binding to the promoter is unknown and can vary depending on the promoter. Marr & Roberts (23) measured a dissociation constant of 3 nM for the $\sigma^{70}$ holoenzyme binding to a 19 bp oligonucleotide containing the TATAAT consensus sequence. We analyze models with a range of values from 1 nM (strong binding) to 1,000 nM (weak binding).

$K_X$. The value of the dissociation constant for the MarA-polymerase complex binding to the promoter is unknown and can vary depending on the promoter. We found reasonable fits by analyzing models in which $K_X$ can be anywhere from 100X smaller to 100X larger than the value of $K_R$.

$[R]_T$. Ishihama (24) and Meuller-Hill (25) estimate the total number of polymerase molecules in the *E. coli* cell at 2,000 and 3,000, respectively. Although *marRAB*, *micF*, and *sodA* are $\sigma^{70}$ promoters, polymerase is distributed among holoenzymes that contain different σ factors in *E. coli*. We used a nominal value of 3,000 copies per cell, and analyzed the sensitivity of the fits to smaller values of 1,000 and 300 copies per cell.

$a_R$ and $a_X$. For a given set of the above parameter values, these parameters are obtained for a given promoter using Eq. (2) by calculating $p_R$ and $p_X$ for values of $[A]_T$ at which measurements are available, and performing a linear regression.

For each model of each promoter, we randomly sampled 10,000 sets of parameter values from the nominal ranges in Table I (Methods) and calculated simulated IPTG-dependent activation profiles. Parameter ranges were sampled in a log uniform manner. After performing the linear regression to calculate values of $a_R$ and $a_X$, fits were evaluated using a standard $\chi^2$ statistic

.

**Acknowledgements**. Supported by funding from the Department of Energy (M.E.W.) and the National Institutes of Health (R.G.M. and J.L.R.). Early modeling and analysis of promoter activity data were made possible by Department of Energy Computational Science Graduate Fellowship Grant DE-FG02-97ER25308 to D.A.M. We thank Jelena Stajic and Luis Bettencourt for assistance in model development, and Ilya Nemenman for comments on the manuscript.

## Figures

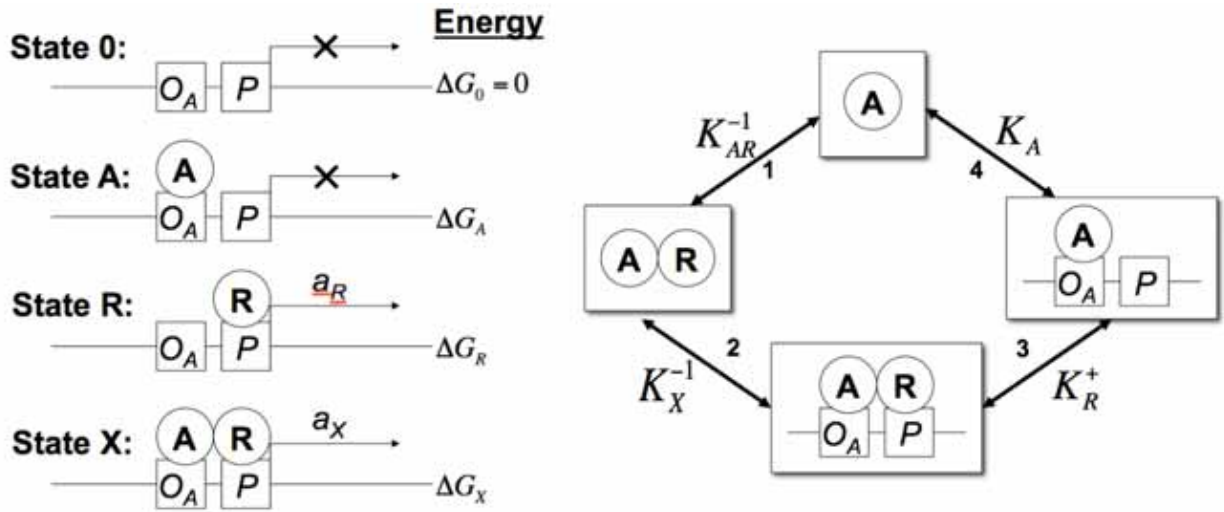

Figure 1. Model of transcriptional activation. *Left panel*. Promoter states and corresponding activities and standard free energies. *Right panel*. Determination of the polymerase-promoter dissociation constant in the presence of MarA ($K_R^+$). The above cycle, followed in a counterclockwise sense from the top, includes: (1) association of MarA and polymerase; (2) association of MarA-polymerase with the promoter; (3) dissociation of polymerase from the promoter when MarA is bound (the process of interest); and (4) dissociation of MarA from the promoter. Using detailed balance, the product of the equilibrium constants for (1)–(4) is equal to 1, yielding $K_R^+ = K_X K_{AR} / K_A$ (Eq. (7)).

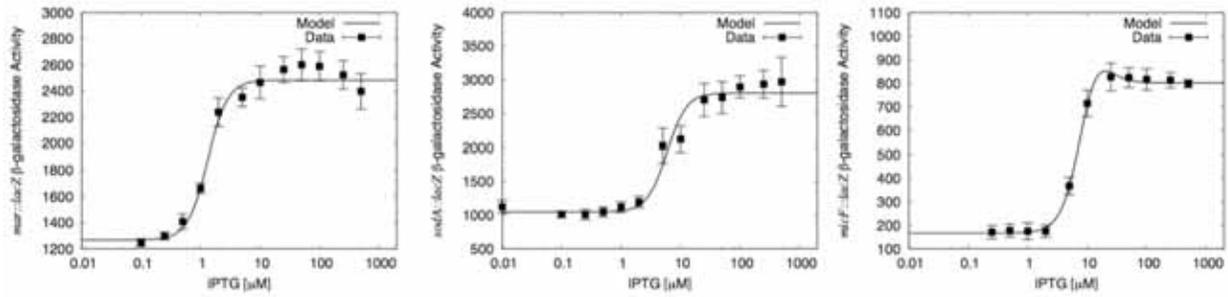

Figure 2: Fit of the best models of *marRAB* activation (*left*) ; *sodA* activation (*center*); and *micF* activation (*right*). Error bars correspond to the standard error of the mean calculated from multiple trials. Corresponding $\chi^2$ and parameter values are given in Table II.

## Tables

Table I. Nominal parameter values used to model activation of *marRAB*, *sodA*, and *micF* promoters by MarA.

| Parameter | *marRAB* | *sodA* | *micF* |
|---|---|---|---|
| $K_{AR}$ [μM] | 21 | 21 | 21 |
| $K_A$ [nM] | 75 | 2,000 | 50 |
| $K_R$ [nM] | (1-1,000) | (1-1,000) | (1-1,000) |
| $K_X$ [nM] | $(0.01\text{-}100)K_R$ | $(0.01\text{-}100)K_R$ | $(0.01\text{-}100)K_R$ |
| $[R]_T$ [Molecules cell$^{-1}$] | 3,000 | 3,000 | 3,000 |

Table II. Properties of models with the lowest value of $\chi^2$. Parameter values were sampled using nominal values and ranges in Table I. Values of $x_{min}$ are listed with asymmetric errors $\sigma_{+x}$ and $\sigma_{-x}$ as $x_{min}(+\sigma_{+x})(-\sigma_{-x})$ (errors are defined in Eq. (9)).

|  | *marRAB* | *sodA* | *micF* |
|---|---|---|---|
| $\chi^2_{min}$ | 6.58 | 7.49 | 0.9 |
| $a_R$ | 1269(+16)(−7) | 1047(+22)(−10) | 167(+7)(−6) |
| $e_a/k_BT$ | −0.678(+0.002)(−0.001) | −1.01(+0.02)(−0.63) | −2.0(+0.3)(−1.9) |
| $e_r/k_BT$ | −0.76(+0.17)(−0.17) | −0.2(+1.2)(−0.5) | +4.5(+2.2)(−0.5) |
| $p^-_{RX}$ | 0.9998(+2×10$^{-5}$)(−0.009) | 0.99985(+9×10$^{-6}$)(−0.015) | 0.9983(+0.001)(−0.0007) |
| $p^+_{RX}/p^-_{RX}$ | 0.9999(+2×10$^{-4}$)(−0.004) | 0.9997(+0.002)(−0.092) | 0.69(+0.2)(−0.3) |

# Model of Transcriptional Activation by MarA in *Escherichia coli*

## *Supporting Information*

Michael E. Wall[1,2,3,*], David A. Markowitz[1], Judah L. Rosner[4], Robert G. Martin[4]

[1] Computer, Computational, and Statistical Sciences Division,

[2] Bioscience Division, and

[3] Center for Nonlinear Studies, Los Alamos National Laboratory, Los Alamos, NM, USA

[4] Laboratory of Molecular Biology, National Institute of Diabetes, Digestive and Kidney Diseases, National Institutes of Health, Bethesda, MD, USA

*Corresponding author

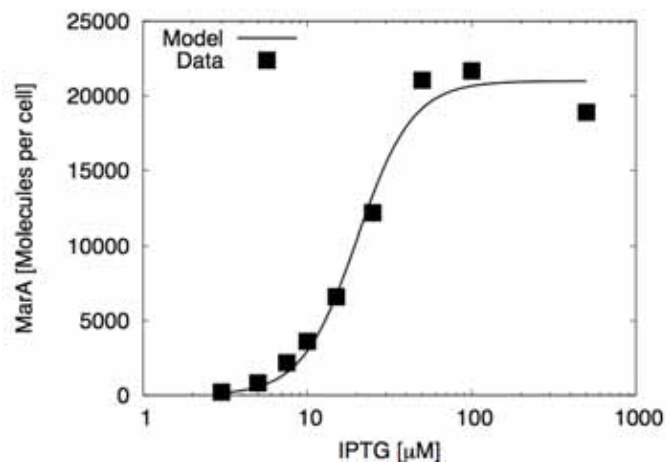

Figure S1: Calibration of IPTG levels against MarA levels. The data (boxes) are well-described by Eq. (5) (line).

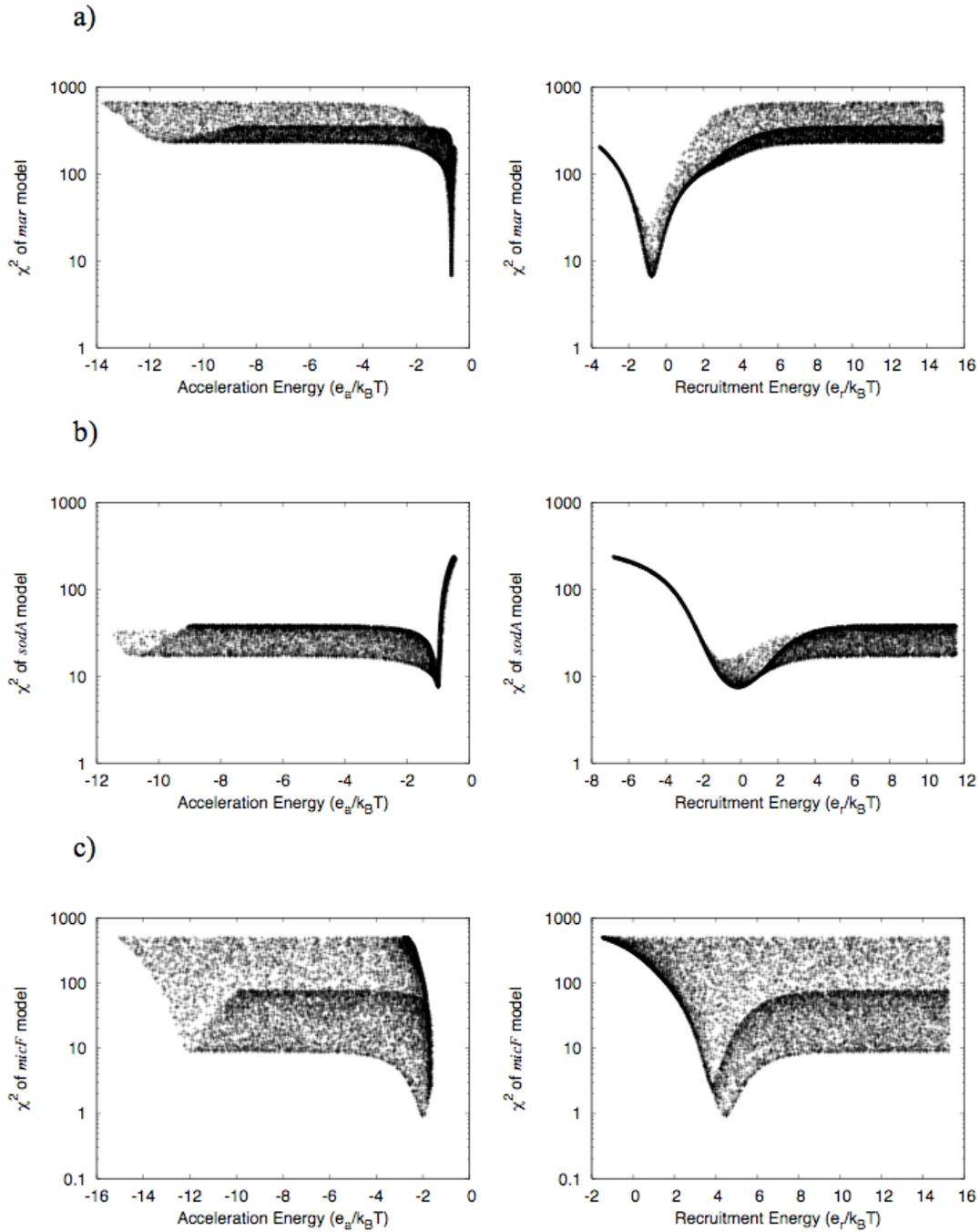

Figure S2. Dependence of $\chi^2$ of a) *marRAB*, b) *sodA*, and c) *micF* models on the acceleration energy (left panels) or attraction energy (right panels). Points correspond to 10,000 different sets of parameter values, sampled using the values in Table I. Points with the lowest $\chi^2$ value correspond to the systems in Table II.

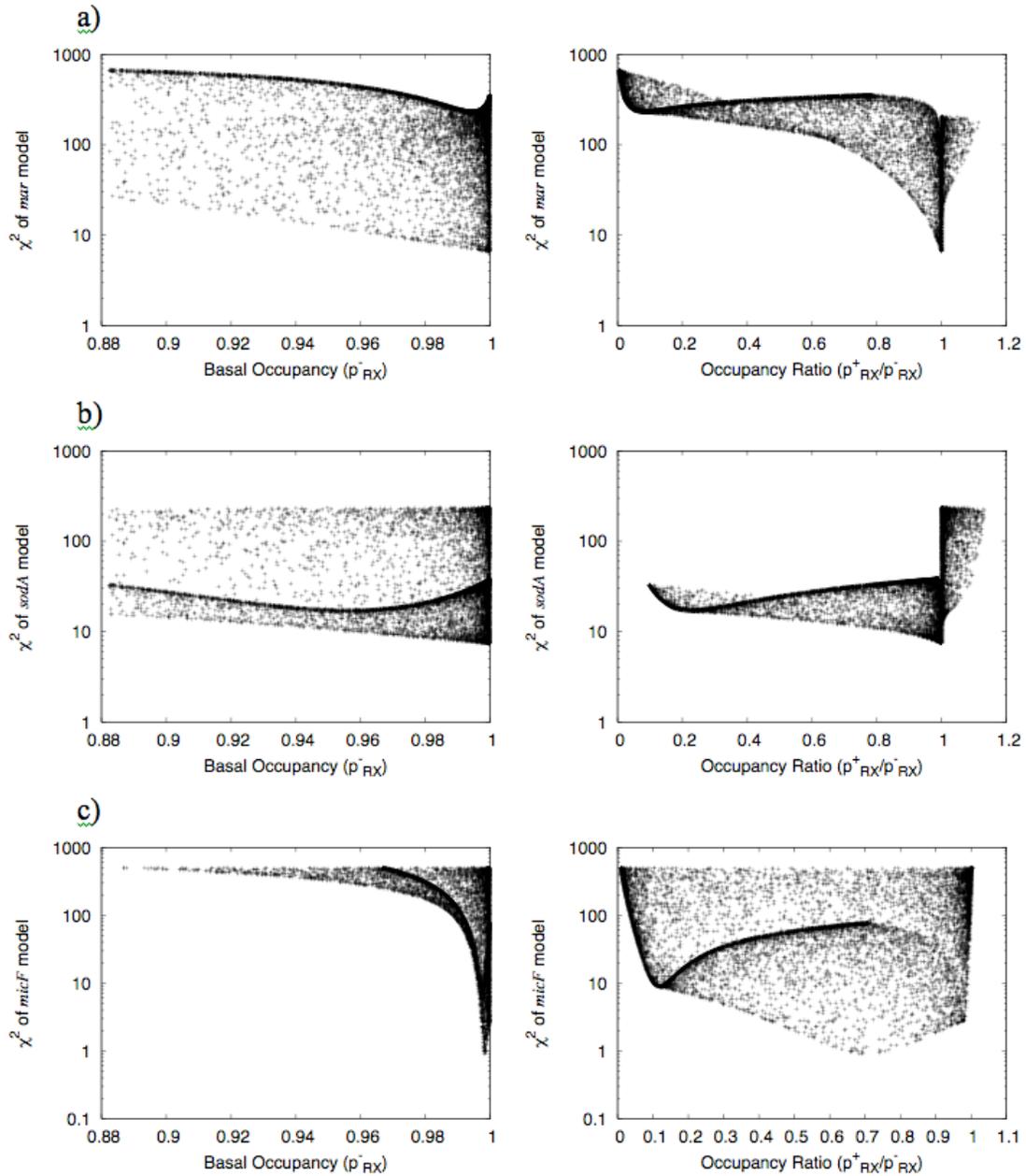

Figure S3. Dependence of $\chi^2$ of a) *mar*, b) *sodA*, and c) *micF* models on the basal occupancy (left panel) or occupancy ratio (right panel). Parameter values were sampled using the nominal values in Table I. For the best models of all promoters, polymerase is bound at the promoter in the absence of activator. For the best models of *mar* and *sodA*, the occupancy remains essentially unchanged in the presence of activator. For the best model of *micF*, the occupancy *decreases* in the presence of activator.

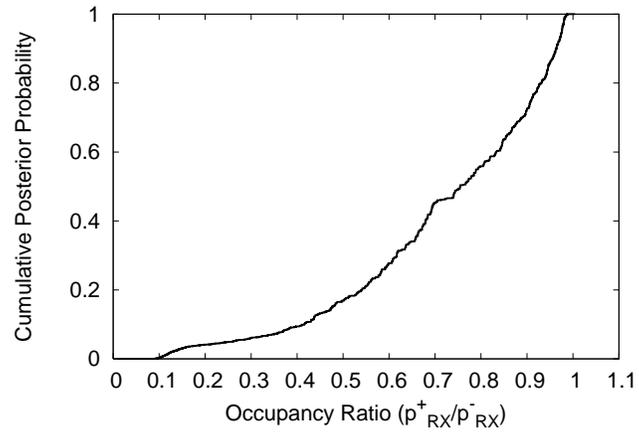

Figure S4. Estimated posterior cumulative probability distribution of the *micF* occupancy ratio. The analysis suggests there is a reasonable chance that the occupancy ratio is measurably less than 1.

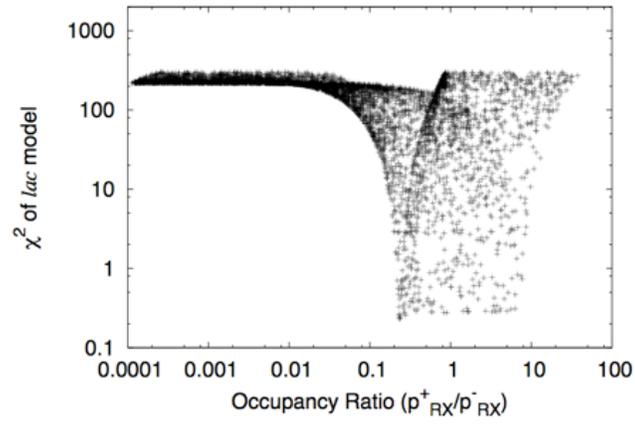

Figure S5: The occupancy ratio can be greater than 1 for low-$\chi^2$ models of *lac* activation by CRP, indicating that recruitment can be a significant factor for this system.

Table S1. Promoter activation data.

| IPTG [μM] | marRAB Mean | Error[a] | sodA Mean | Error[a] | micF Mean | Error[a] |
|---|---|---|---|---|---|---|
| 0 | | | 1124 | 98 | | |
| 0.1 | 1247 | 29 | 1012 | 45 | | |
| 0.25 | 1298 | 28 | 1010 | 79 | 171 | 29 |
| 0.5 | 1409 | 57 | 1050 | 74 | 177 | 28 |
| 1 | 1663 | 40 | 1119 | 79 | 175 | 35 |
| 2 | 2238 | 109 | 1196 | 84 | 175 | 26 |
| 5 | 2352 | 72 | 2025 | 256 | 367 | 37 |
| 10 | 2466 | 125 | 2120 | 194 | 714 | 57 |
| 25 | 2565 | 98 | 2701 | 248 | 826 | 59 |
| 50 | 2601 | 122 | 2737 | 237 | 824 | 42 |
| 100 | 2590 | 112 | 2895 | 165 | 816 | 45 |
| 250 | 2524 | 110 | 2936 | 207 | 813 | 32 |
| 500 | 2397 | 135 | 2970 | 365 | 798 | 17 |

[a]Standard error of the mean $\sigma/\sqrt{N}$ calculated from $N$ measurements.